\documentclass[12pt, reqno]{amsart}
\usepackage{amsfonts,latexsym,enumerate}
\usepackage{amsmath}
\usepackage{amscd}
\usepackage{float,amsmath,amssymb,mathrsfs,bm,multirow,graphics}
\usepackage[dvips]{graphicx}
\usepackage[percent]{overpic}
\usepackage[pdftex]{color}
\usepackage[numbers,sort&compress]{natbib}
\usepackage{amsaddr}

\newcommand{\nequation}{\setcounter{equation}{0}}

\newcommand{\R}{{\Bbb R}}

\newcommand{\C}{{\Bbb C}}

\newcommand{\proofend}{\hfill$\Box$\bigskip}
\newcommand{\proofendcontinue}{\hfill \raisebox{.8mm}[0cm][0cm]{$\bigtriangledown$}\bigskip}

\newcommand{\im}{\text{\upshape Im\,}}

\def\XXint#1#2#3{{\setbox0=\hbox{$#1{#2#3}{\int}$}
\vcenter{\hbox{$#2#3$}}\kern-.5\wd0}}

\newtheorem{theorem}{Theorem}

\newtheorem{remark}{Remark}

\input epsf
\title[Absence of solitons for the defocusing NLS]
{Absence of solitons for the defocusing NLS equation on the half-line}

\author{Jonatan Lenells}
\address{Department of Mathematics, KTH Royal Institute of Technology, \\ 100 44 Stockholm, Sweden.}
\email{jlenells@kth.se}

\begin{document}
\begin{abstract} 
\noindent
It has been conjectured that the defocusing nonlinear Schr\"odinger (NLS) equation on the half-line does not admit solitons. We give a proof of this conjecture.
\end{abstract}

\maketitle




\section{Introduction}\nequation
The long time behavior of the solution of the nonlinear Schr\"odinger (NLS) equation
 on the line can be determined via inverse scattering techniques and the nonlinear steepest descent method \cite{DIZ1993, I1981, M1974}. The asymptotic solitons are generated by the zeros in the upper half plane of a certain spectral function. Under the assumption that the solution vanishes as $|x| \to \infty$, this function can have zeros in the focusing case whereas it cannot have zeros in the defocusing case, corresponding to the fact that the focusing NLS admits solitons whereas the defocusing NLS does not. 

It has been conjectured that solitons are absent also for the defocusing NLS on the half-line \cite{FIS2005, MQ2014}. This is a natural conjecture in view of the situation on the line, but it is not immediate that it holds: for example, it could be that there are soliton-like structures which are singular on the line, but which become regular when restricted to the half-line. 
For the half-line problem, the asymptotic solitons for the NLS equation are generated by the zeros in the second quadrant of a certain spectral function $d(k)$ \cite{FIS2005}. In this note, we use an approximation argument together with Rouch\'e's theorem to show that $d(k)$Ê has no zeros in the second quadrant in the defocusing case. This establishes the absence of solitons for the defocusing NLS on the half-line.
 
Let us point out that although the defocusing NLS equation does not admit solitons that vanish at infinity, it does admit soliton solutions which have a nontrivial background intensity, sometimes referred to as dark solitons, see \cite{FT2007, KL1998}. In this note, we only consider solutions that decay as $x \to \infty$. 

Finally, we mention that the formalism of \cite{FIS2005}, and hence also our result here, only applies when the Dirichlet and Neumann boundary values of the solution decay for large $t$. It has been shown, at least in the case of vanishing initial data, that if the Dirichlet data decays as $t \to \infty$, then so does the Neumann value \cite{AK2014}.

\section{Main result}
Consider the defocusing nonlinear Schr\"odinger (NLS) equation
 \begin{align}\label{nls}
  iq_t + q_{xx} - 2 |q|^2 q = 0,
\end{align}
in the quarter plane domain $\{x > 0, t> 0\}$.\footnote{Carroll and Bu showed in \cite{CB1991} that the Dirichlet problem for (\ref{nls}) is well-posed: Given initial data in $H^2$ and compatible Dirichlet data in $C^2$, there exists a unique global solution $u \in C^1(L^2) \cap C^0(H^2)$.}
Let $\R_+ = (0,\infty)$. We assume that $q(x,t)$ is a sufficiently smooth solution such that $q(x,t)$ decays as $x \to \infty$ for each $t \geq 0$ and such that the Dirichlet and Neumann values, denoted by  $g_0(t) = q(0,t)$ Êand $g_1(t) = q_x(0,t)$, decay as $t \to \infty$; for example, the following decay assumptions on $\{g_j(t)\}$ are sufficient:
\begin{align}\label{gjdecay}
(1 + t)\partial^ig_j(t) \in L^1(\R_+), \qquad j = 0,1, \quad i = 0,1,2.
\end{align}

Under the above assumptions, the formalism of \cite{FIS2005} represents the solution of (\ref{nls}) in terms of the solution of a $2 \times 2$-matrix valued Riemann-Hilbert (RH) problem, whose formulation involves four spectral functions $\{a(k), b(k), A(k), B(k)\}$ defined as follows: 
$$s(k) := \begin{pmatrix} \overline{a(\bar{k})} & b(k) \\ \overline{b(\bar{k})} & a(k) \end{pmatrix} = \mu_3(0,0,k), \qquad 
S(k) := \begin{pmatrix} \overline{A(\bar{k})} & B(k) \\ \overline{B(\bar{k})} & A(k) \end{pmatrix} = \mu_1(0,0,k),$$
where the eigenfunctions $\{\mu_j(x,t,k)\}_1^3$ satisfy
\begin{align}\label{muVolterra}
\mu_j(x,t,k) = I + \int_{(x_j, t_j)}^{(x,t)} e^{-i(k(x - x') + 2k^2(t - t'))\hat{\sigma}_3}(Q\mu_j dx' + \tilde{Q} \mu_jdt')
\end{align}
with $(x_1, t_1) = (0,\infty)$, $(x_2, t_2) = (0,0)$, $(x_3,t_3) = (\infty, t)$, and
$$Q(x,t) = \begin{pmatrix} 0 & q \\ \bar{q} & 0 \end{pmatrix}, \qquad \tilde{Q}(x,t,k) = 2kQ - iQ_x \sigma_3 - i|q|^2\sigma_3, \qquad \sigma_3 = \begin{pmatrix} 1 & 0 \\ 0 & -1 \end{pmatrix}.$$ 

We define $c(k)$ and $d(k)$ for $k \in \R \cup i\R_+$ by
\begin{align}\label{cddef}
\begin{pmatrix} \overline{d(\bar{k})} & c(k) \\ \overline{c(\bar{k})} & d(k) \end{pmatrix} = S^{-1}(k)s(k).
\end{align}
The expressions
$$c(k) = A(k)b(k) - B(k)a(k), \qquad
d(k) = a(k)\overline{A(\bar{k})} -  b(k) \overline{B(\bar{k})},$$
show that $c(k)$ and $d(k)$ have bounded analytic continuations to $D_1$ and $D_2$, respectively, where $D_j = \{(j-1)\pi/2 < \arg k < \pi/2\}$, $j = 1, \dots, 4$, denote the four quadrants of the complex $k$-plane.
An application of the nonlinear steepest descent method of \cite{DZ1993} reveals that the asymptotic solitons are generated by the zeros in $D_2$ of the function $d(k)$, see Theorem B.1 in \cite{FIS2005}. 
We prove the following result.

\begin{theorem}\label{mainth}
 The function $d(k)$ has no zeros in $\bar{D}_2$.
\end{theorem}

\section{Proof of Theorem \ref{mainth}}
Let $\mu_1^{(T)}(x,t,k)$ denote the solution of (\ref{muVolterra}) normalized at $(x, t) = (0,T)$, i.e. $\mu_1^{(T)}$ satisfies (\ref{muVolterra}) with $(x_j, t_j) = (0,T)$. 
Let $S(T,k) = \mu_1^{(T)}(0,0,k)$.
In analogy with (\ref{cddef}), we define functions $c(T,k)$ and $d(T,k)$ for $\im k \geq 0$ by
$$\begin{pmatrix}Ê\overline{d(T, \bar{k})} & c(T,k) \\ \overline{c(T,\bar{k})} & d(T,k) \end{pmatrix}
= S(T,k)^{-1}s(k) = e^{2ik^2T\hat{\sigma}_3} \mu_3(0,T,k).$$

\bigskip\noindent
{\bf Claim 1.} $a(k) \neq 0$ for every $k \in \bar{D}_1 \cup \bar{D}_2$.
\medskip

\noindent
{\it Proof of Claim 1.}
The determinant relation
$$1 = \det s(k) = |a(k)|^2 - |b(k)|^2, \qquad k \in \R,$$
implies that $a(k)$ is nonzero for $k \in \R$.	
Suppose $a(\kappa) = 0$ for some $\kappa \in \C$ with $\im \kappa > 0$. Consider the space $L^2(\R, \C^2)$ of vector valued functions $f = (f_1, f_2)$ equipped with the inner product
$$\langle f, g\rangle = \int_\R (\bar{f}_1 g_1 + \bar{f}_2 g_2) dx.$$
Let 
$$q_e(x) = \begin{cases} q(x, 0), & x \geq 0, \\ 0, & x < 0, \end{cases} \quad \text{and} \quad Q_e = \begin{pmatrix} 0 & q_e \\ \bar{q}_e & 0 \end{pmatrix}.$$
Then the operator $L = i\sigma_3 \partial_x - i\sigma_3Q_e$ satisfies
$$\langle Lf, g \rangle = \langle f, Lg\rangle \quad \text{whenever} \quad f,g \in H^1(\R, \C^2) \subset L^2(\R, \C^2).$$
Define $h \in L^2(\R, \C^2)$ by
$$h(x) = \begin{cases}
[\mu_3(x,0,\kappa)]_2 e^{i\kappa x}, & x \geq 0, \\
\begin{pmatrix} b(\kappa) e^{-i\kappa x}   \\ 0 \end{pmatrix}, & x < 0.
\end{cases}$$
The assumption $a(\kappa) = 0$ implies that $h$ is continuous at $x = 0$. Moreover, since $\im \kappa > 0$, $h$ has exponential decay as $x \to \pm \infty$. It follows that $h \in H^1(\R, \C^2)$. But since $Lh = \kappa h$ this leads to the contradiction that the eigenvalue $\kappa$ must be real:
$$\bar{\kappa} \langle h, h \rangle = \langle Lh, h \rangle =\langle h, Lh \rangle = \kappa \langle h, h \rangle.$$
This proves the claim.
\proofendcontinue

\bigskip\noindent
{\bf Claim 2.} $d(k) \neq 0$ for every $k \in \R \cup i\R_+$.
\medskip

\noindent
{\it Proof of Claim 2.}
The determinant relation
$$1 = \det[S(k)^{-1}s(k)] = |d(k)|^2 - |c(k)|^2, \qquad k \in \R,$$
implies that $d(k)$ is nonzero for $k \in \R$. 
On the other hand, the initial and boundary values satisfy the following so-called global relation, see equation (3.19) in \cite{FIS2005}:
\begin{align}\label{GR}
c(k) = 0, \qquad k \in \bar{D}_1.
\end{align}
The global relation (\ref{GR}) together with the determinant relation $\det S(k) = 1$ imply that
$$d(k) = a(k)\overline{A(\bar{k})} - \frac{B(k)a(k)}{A(k)}\overline{B(\bar{k})} = \frac{a(k)}{A(k)}, \qquad k \in i\R_+.$$
Since $A(k)$ is bounded in $\bar{D}_1$ and $a(k)$ is nonzero in the upper half plane by Claim 1, it follows that $d(k)$ is nonzero on $i\R_+$.
\proofendcontinue

\bigskip\noindent
{\bf Claim 3.} There exists a constant $C > 0$ such that
\begin{align}\label{ddiffbound}
|d(T,k) - d(k)| \leq \frac{C}{1 + T}, \qquad T \geq 0, \quad k \in \partial D_2.
\end{align}
\medskip

\noindent
{\it Proof of Claim 3.}
The $(22)$ element of the relation
$$S(T,k)^{-1}s(k) = [e^{2ik^2T\hat{\sigma}_3}(\mu_1(0, T,k))] S(k)^{-1} s(k), \qquad k \in \R \cup i\R_+,$$
implies that
\begin{align}\label{dTk}
d(T,k)  = e^{-4ik^2T}(\mu_1(0, T,k))_{21} c(k) + (\mu_1(0,T,k))_{22} d(k), \qquad k \in \R \cup i\R_+.
\end{align}
Hence, for $k \in \R \cup i\R_+$,
\begin{align}\label{ddiff}
|d(T,k) - d(k)| \leq |(\mu_1(0, T,k))_{21}| |c(k)| + |(\mu_1(0, T,k))_{22} - 1| |d(k)|.
\end{align}
The decay assumption (\ref{gjdecay}) implies that 
\begin{align}\label{mu1est}
\mu_1(0,T,k) = I + O(T^{-1}), \qquad T \to \infty, \quad k \in \R \cup i\R,
\end{align}
where the error term is uniform with respect to $k$ in the given range. 
Equation (\ref{ddiffbound}) follows from (\ref{ddiff}) and (\ref{mu1est}). 
\proofendcontinue

\bigskip\noindent
{\bf Claim 4.} There exists a $T_0 > 0$ such that $d(T,k) \neq 0$ for each $k \in \bar{D}_2$ and each $T \geq T_0$.
\medskip

\noindent
{\it Proof of Claim 4.}
By Claims 2 and 3, there exists a $T_0 > 0$ such that $d(T,k) \neq 0$ for $k \in \partial D_2$ and $T \geq T_0$. Fix $T \geq T_0$ and suppose $d(T,\kappa) = 0$ for some $\kappa \in D_2$. 
We let $L = i\sigma_3 \partial_x - i\sigma_3Q_e$, where
$$q_e(x) = \begin{cases} q(x,T), & x \geq 0, \\ 0, & x < 0, \end{cases} \quad \text{and} \quad Q_e = \begin{pmatrix} 0 & q_e \\ \bar{q}_e & 0 \end{pmatrix},$$
and define $h \in L^2(\R, \C^2)$ by
$$h(x) = \begin{cases}
[\mu_3(x, T, \kappa)]_2 e^{i\kappa x + 2i\kappa^2T}, & x \geq 0, \\
\begin{pmatrix} c(T, \kappa) e^{-i\kappa x - 2i\kappa^2 T}  \\ 0 \end{pmatrix}, & x < 0.
\end{cases}$$
The condition $d(T, \kappa) = 0$ implies that $h$ is continuous at $x = 0$. As in the proof of Claim 1, the facts that $h \in H^1(\R, \C^2)$ and $Lh = \kappa h$ lead to the contradiction that $\kappa \in \R$.
\proofendcontinue

\bigskip\noindent
{\bf Claim 5.} $d(k) \neq 0$ for every $k \in \bar{D}_2$.
\medskip

\noindent
{\it Proof of Claim 5.}
Since $d(k) = 1 + O(k^{-1})$ and $d(T, k) = 1 + O(k^{-1})$ as $k \to \infty$ in $\bar{D}_2$, the inequality
\begin{align}\label{rouche}  
  |d(T,k) - d(k)| < |d(k)|
\end{align}
holds for all sufficiently large $k \in \bar{D}_2$. Claim 2 shows that $d(k)$ has no zeros in $\partial D_2$.  Hence, by Claim 3, we can choose $T \geq T_0$ so that (\ref{rouche}) holds also on $\partial D_2$. Rouch\'e's theorem then implies that $d(T, k)$ and $d(k)$ have the same number of zeros in $D_2$. Since $d(T,k)$ has no zeros in $\bar{D}_2$ by Claim 4, $d(k)$ also has no zeros in $\bar{D}_2$. This completes the proof of the theorem.
\proofend

\begin{remark}\upshape
Rouch\'e's theorem is usually stated under the assumption that the functions are analytic in a domain containing the given contour. The function $d(k)$ is analytic in $D_2$ but only continuous on $\bar{D}_2$ in general. This technical issue can be circumvented, for example, by applying Rouch\'e's theorem to a sequence of closed contours $C_n \subset D_2$ converging to $\partial D_2$ as $n \to \infty$, noting that the inequality (\ref{rouche}) holds also near $\partial D_2$ by continuity.
\end{remark}

\bigskip
\noindent
{\bf Acknowledgement} {\it The author is grateful to Peter D. Miller for valuable suggestions and acknowledges support from the EPSRC, UK.}

\bibliographystyle{plain}
\bibliography{is}

\end{document}